# MASSES AND AGES FOR 230,000 LAMOST GIANTS, VIA THEIR CARBON AND NITROGEN ABUNDANCES

Anna Y. Q. Ho,[1,2] Hans-Walter Rix,[2] Melissa K. Ness,[2]
David W. Hogg,[2,3,4,5] Chao Liu,[6] and Yuan-Sen Ting (丁源森)[7,8]

[1]*Cahill Center for Astrophysics, California Institute of Technology, MC 249-17, 1200 E California Boulevard, Pasadena, CA, 91125, USA*

[2]*Max-Planck-Institut für Astronomie, Königstuhl 17, D-69117 Heidelberg, Germany*

[3]*Simons Center for Data Analysis, 160 Fifth Avenue, 7th Floor, New York, NY 10010, USA*

[4]*Center for Cosmology and Particle Physics, Department of Phyics, New York University, 4 Washington Place, Room 424, New York, NY, 10003, USA*

[5]*Center for Data Science, New York University, 726 Broadway, 7th floor, New York, NY 10003, USA*

[6]*Key Laboratory of Optical Astronomy, National Astronomical Observatories, Chinese Academy of Sciences, Datun Road 20A, Beijing 100012, China*

[7]*Harvard Smithsonian Center for Astrophysics, 60 Garden Street, Cambridge, MA 02138, USA*

[8]*Research School of Astronomy and Astrophysics, Mount Stromlo Observatory, Australian National University, Cotter Road, Weston Creek, ACT 2611, Australia*

## ABSTRACT

We measure carbon and nitrogen abundances to a precision of $\lesssim 0.1$ dex for $450,000$ giant stars from their low-resolution ($R \sim 1800$) LAMOST DR2 survey spectra. We use these [C/M] and [N/M] measurements, together with empirical relations based on the APOKASC sample, to infer stellar masses and implied ages for 230,000 of these objects to 0.08 dex and 0.2 dex respectively. We use *The Cannon*, a data-driven approach to spectral modeling, to construct a predictive model for LAMOST spectra. Our reference set comprises 8125 stars observed in common between the APOGEE and LAMOST surveys, taking seven APOGEE DR12 labels (parameters) as ground truth: $T_{\rm eff}$, $\log g$, [M/H], [$\alpha$/M], [C/M], [N/M], and $A_{\rm k}$. We add seven colors to the Cannon model, based on the *g, r, i, J, H, K, W1, W2* magnitudes from APASS, 2MASS, and *WISE*, which improves our constraints on $T_{\rm eff}$ and $\log g$ by up to 20% and on $A_{\rm k}$ by up to 70%. Cross-validation of the model demonstrates that, for high-S/N objects, our inferred labels agree with the APOGEE values to within 50 K in temperature, 0.04 mag in $A_{\rm k}$, and $< 0.1$ dex in $\log g$, [M/H], [C/M], [N/M], and [$\alpha$/M]. We apply the model to 450,000 giants in LAMOST DR2 that have not been observed

ah@astro.caltech.edu

by APOGEE. This demonstrates that precise individual abundances can be measured from low-resolution spectra and represents the largest catalog to date of homogeneous stellar [C/M], [N/M], masses, and ages. As a result, we greatly increase the number and sky coverage of stars with mass and age estimates.

1. INTRODUCTION

An empirical description of the Milky Way's present structure and formation history requires accurate and consistent age estimates for large samples of stars distributed throughout the galaxy. Although we have recently entered an era of extensive spatial, kinematic, and chemical information beyond the solar neighborhood, comparably extensive age constraints remain elusive.

Stellar age is a property that must be inferred from observations with the help of stellar evolution models; generally, it cannot be measured "directly." Therefore, results are inherently limited by the applicability and accuracy of the model used (see Soderblom 2010 for a comprehensive review). Because stellar ages are difficult to measure directly, abundances such as [Fe/H] and [$\alpha$/Fe] are commonly used as an age-dating proxy (e.g. via making maps of mono-age populations; see Rix & Bovy (2013) and Bovy et al. (2016)) because the determination of photospheric abundances from spectra is more straightforward.

Unfortunately for Milky Way studies, the population of stars that is most readily observable throughout the galaxy — red giant stars — is also the one for which it is particularly challenging to estimate ages. The standard technique of isochrone fitting is impractical in this regime, because there is too much uncertainty both in stellar parameter measurements and in the model isochrones. In other words, stars with very different ages can have very similar atmospheric parameters and luminosities (Soderblom 2010; Rix & Bovy 2013).

Instead, the key to age-dating red giant stars is mass. Because the red giant phase is so short, the age of a star is essentially its main-sequence lifetime, which is set by the star's mass and metallicity (Soderblom 2010). Given mass and metallicity, one can estimate age using isochrones, e.g. by interpolating between them.

Recently, asteroseismology has made it possible to measure masses for giants out to large distances. The *Kepler* mission (Bedding et al. 2010; Borucki et al. 2010; Koch et al. 2010; Gilliland et al. 2011) has conducted long-cadence photometry for over 10,000 giants along a pencil-beam through the Galaxy (Stello et al. 2013). From detailed light curves, one can measure two characteristic variability frequencies that directly probe the (age-dependent) structure of the stellar interior: $\nu_{\max}$ is the frequency corresponding to the maximum oscillation power, and $\Delta\nu$ is the frequency spacing between two consecutive modes of the same spherical degree. This approach is especially effective for giants because they have higher densities and thus a larger sound speed, which makes these (acoustic) oscillations more pronounced (Soderblom 2010). Together with a measurement of the star's $T_{\text{eff}}$, and the solar values $\nu_{\max,\odot}$ and $\Delta\nu_\odot$, the mass of the star can be estimated using seismic scaling relations. Note that these scaling relations are based on Sun-like stars, and may not be suitable for metal-poor stars (Epstein et al. 2014).

Furthermore, the population of stars with asteroseismic measurements is spatially limited. Ness et al. (2016) and Martig et al. (2016) greatly expanded the spatial coverage of giants with age estimates by determining masses spectroscopically: they showed that the masses (and implied ages) of post dredge-up giants can be measured from high-resolution infrared (APOGEE, $R \approx 22,500$) spectra, and determined a model of mass and age as a function of $T_{\rm eff}$, $\log g$, [M/H], [C/M], and [N/M] values (see Tables A2 and A3 in Martig et al. 2016). Their work increased the sample of giant stars with known ages to 70,000, the largest (and most spatially extended) sample of stellar ages to date.

To make these spectroscopic mass measurements, Ness et al. (2016) used a data-driven approach to stellar modeling called *The Cannon* (Ness et al. 2015). For a complete description of the methodology, see Ness et al. (2015); we summarize briefly here. In short, *The Cannon* can be used to transfer a system of stellar "labels" (physical parameters and element abundances) from one survey to another, via a training step and a test step. In the training step, *The Cannon* uses a reference set of objects observed in common between the surveys of interest to construct a predictive model of spectra independently at each wavelength. For example, for a set of objects measured in common between Survey A and Survey B, *The Cannon* might fit a model that predicts every pixel in a Survey A spectrum given Survey B labels. In the test step, this model can be used to infer new labels directly from Survey A spectra that are by construction on the Survey B label scale. *The Cannon* uses no explicit physical stellar models, is very fast, and achieves comparable accuracy to existing survey pipelines using spectra of significantly lower S/N; it requires only a set of objects observed in common between the surveys.

Via this "label transfer," *The Cannon* has shown promise for bringing qualitatively different stellar surveys onto a consistent physical scale. For example, Ho et al. (2017) used *The Cannon* to transfer labels from a high-resolution, high-S/N survey (APOGEE) to a low-resolution, modest-S/N survey (LAMOST). They showed that basic parameters ($T_{\rm eff}$, $\log g$, [Fe/H], and [$\alpha$/M]) consistent with APOGEE values can be determined directly from LAMOST spectra, using *The Cannon*. This enabled the largest and most spatially extended sample of [$\alpha$/M] values to date ($\sim$450,000 giants).

Taken together, the work in Martig et al. (2016), Ness et al. (2016), and Ho et al. (2017) raises the prospect that masses and ages could be determined for a large sample of optical giant spectra. After all, in APOGEE spectra (the near-IR), mass is encoded in CN and CO molecular regions, and [C/N] and [C/H] features are prominent in the blue regions ($\sim 4100$Å) of giant spectra (e.g. Martell et al. 2008). Recent theoretical work by Salaris et al. (2015) and Martig et al. (2016) lends physical justification to why these features should be indicative of mass: during a star's main-sequence lifetime, the CNO cycle in its core determines the final relative abundances of carbon and nitrogen. After arriving on the giant branch, the material in the core is dredged up to the surface via convective mixing. The depth of the convective envelope, and the [C/N] ratio in the core, is determined by the mass of the star. Thus, in giants that have undergone dredge-up once (that is, they have not undergone additional convective mixing) the [C/N] ratio observed in the photosphere is (together with metallicity) highly predictive of mass.

For a large set of stellar masses and ages, we turn to the Large Sky Area Multi-Object Fiber Spectroscopic Telescope (LAMOST), the largest ongoing stellar spectroscopic survey. We measure mass and age by using *The Cannon* to transfer a label system from APOGEE, as in Ho et al. (2017). Our work extends Ho et al. (2017) by two additional labels ([C/M] and [N/M]). We follow that paper very closely and refer to it from now on as Paper 1.

The challenge in measuring mass and age from LAMOST spectra via [C/M] and [N/M] is that [C/M] and [N/M] have not traditionally been measured from low-resolution ($R \lesssim 5000$) spectra. A data-driven approach, however, will enable us to measure [C/M] and [N/M] from these spectra to $\lesssim 0.1\,\mathrm{dex}$. From there, we use the theoretical relations in Martig et al. (2016) to determine masses and ages for as many giant stars as possible, restricted primarily by the parameter regime in which the relations are applicable. This will enable us to measure the largest sample of stellar ages to date.

In Section 2 we briefly recapitulate the data sets. In Section 3 we recapitulate the mathematics of the label transfer, from which we derive carbon and nitrogen abundances in Section 4 and ages and masses in Section 5.

## 2. THE LAMOST AND APGOGEE SURVEYS

Here we give a very brief summary of the two spectroscopic surveys underlying the analysis in this paper.

LAMOST is a low-resolution ($R \approx 1,800$) optical ($3650 - 9000\,\mathrm{\AA}$) moderate-S/N survey. The second data release (DR2; Luo et al. 2015) was recently made public and includes spectra and three parameters ($T_{\mathrm{eff}}$, $\log g$, [Fe/H]) for 2.2 million stars, of which 500,000 are giants (Liu et al. 2014). These parameters are measured by the LAMOST Stellar Parameter pipeline (LASP; Wu et al. 2011a,b; Luo et al. 2015) which uses the Correlation Function Initial (CFI; Du et al. 2012) to determine an initial coarse solution, and the Université de Lyon Spectroscopic Analysis Software (*ULySS*; Koleva et al. 2009; Wu et al. 2011b) to determine the final solution. The grid of model spectra used by *ULySS* come from the ELODIE spectral library (Prugniel & Soubiran 2001; Prugniel et al. 2007).

By contrast, APOGEE (the Apache Point Observatory Galactic Evolution Experiment; Majewski et al. 2015) is a high-resolution ($R \approx 22,500$), high-S/N (S/N $\approx 100$), H-band (15200-16900 Å) survey of giant stars in the Milky Way bulge, disk, and halo, part of the Sloan Digital Sky Survey III (Eisenstein et al. 2011). Observations are conducted using a 300-fiber spectrograph (Wilson et al. 2010) on the 2.5 m Sloan Telescope (Gunn et al. 2006) at the Apache Point Observatory (APO) in Sunspot, New Mexico (USA). Data Release 12 (DR12; Alam et al. 2015; Holtzman et al. 2015) comprises spectra for >100,000 giants together with stellar parameters and 15 chemical abundances measured by the ASPCAP pipeline (García Pérez et al. 2016) and FERRE code (Allende Prieto et al. 2006). Thus, LAMOST represents a large expansion over APOGEE not only in sample size, but also in area coverage (LAMOST stars are measured away from the disk, unlike APOGEE) and parameter range (in particular, [Fe/H]).

## 3. LABEL TRANSFER USING *The Cannon*

As described in Section 1, there are two steps to *The Cannon*: a training step, in which the coefficients of the spectral model are fit for, and a test step, in which the spectral model is used to infer labels from new spectra. Our implementation of these steps for transferring a label system from APOGEE to LAMOST closely follows Paper 1. Here, too, the data consist of spectra from LAMOST, which we again normalize in a consistent way using a running Gaussian, and labels from APOGEE DR12. As before, the spectral model is quadratic in the labels. After training for and fixing the model, we apply it to measure labels for the 454,180 giants from Paper 1, identified via the procedure in Section 4.2 of that paper. Despite the fact that we follow Paper 1 very closely, there are important differences in (and new components to) our modeling.

### 3.1. *Differences in the Modeling Procedure from Paper 1*

There are important differences in (and new components to) our modeling as compared to Paper 1: in the labels that we use, in the reference objects that we use to train the model, in spectral regions that we mask out, and in the incorporation of photometry.

Here, our model is quadratic in seven labels instead of the original five labels: we use $T_{\text{eff}}$, $\log g$, [M/H], [C/M], [N/M], [$\alpha$/M], and K-band extinction $A_k$. Because we will eventually use the relations in Martig et al. (2016) to translate our carbon and nitrogen abundances into age estimates, our labels need to be on the same scale as those that were used to fit for the relations. Thus, whereas we used the calibrated DR12 parameters in Paper 1 (those in the `PARAM` array), in this case, we use the raw, uncalibrated values from the `FPARAM` array.

Furthermore, we do not use the full reference set of 9956 objects from Paper 1, because some of these have unreliable [C/M] and [N/M] reference labels. Following Martig et al. (2016), we excise the 743 objects that have both $T_{\text{eff}} > 4550$ and $-1 <$ [M/H] $< -0.5$ in order to eliminate objects with only an upper limit measurement on [C/M] and a lower limit on [N/M]. In addition, Martig et al. (2016) found that the minimum [C/M] possible to measure is on the level of -0.4 to -0.5 dex, so we also exclude the 40 objects with [C/M] $< -0.4$. This left us with 9173 out of the original 9956.

In Paper 1, we fit an independent spectral model at every spectral pixel. However, there are features in LAMOST spectra that arise from effects in a different velocity system from that of the star: for example, diffuse interstellar bands (DIBs) and the Na

I doublet are interstellar absorption features originating from intervening material. We noticed by examining the leading coefficients of an initial Cannon model (see Figure 3) that *The Cannon* was "learning" to use these features to predict labels intrinsic to the star, particularly [$\alpha$/M], and that this introduced radial velocity-dependent systematic errors into the label estimates[1]. The leading coefficients of an initial Cannon model also indicated that the imperfectly corrected telluric bands, originating in the Earth's atmosphere, left small, but significant, velocity-dependent effects in the rest-frame spectra of the stars.

To prevent *The Cannon* from using these features spuriously, we masked them out by setting the inverse variances corresponding to these pixels to be zero. To be conservative, roughly half of each spectrum was masked, but because the most important features for our labels of interest were preserved, this still improved our label estimates.

Masking out the interstellar absorption features took out most of the spectral information on $A_k$, which originated in the DIB strength. However, of course, multi-band photometry of a star encodes a combination of its effective temperature and its reddening. We found that incorporating photometry not only enabled us to accurately and precisely predict $A_k$ for our reference objects, but also improved the precision of our estimate of $T_{eff}$, particularly for lower-S/N spectra. (Note that we use the term "accurate" to mean unbiased, and the term "precise" to mean low variance.) Incorporating photometry also improved the precision of our estimate of $\log g$, presumably because measurements of $T_{eff}$ and $\log g$ are highly covariant for spectra of this resolution due to blending (see, e.g. Ting et al. (2017); submitted to ApJ). The scatter in $A_k$ decreased by 70% in the lowest-S/N spectra, by 50% in spectra with $50 < S/N < 100$, and by 25% in spectra with $S/N > 100$. For low-S/N spectra, the scatter in $T_{eff}$ and $\log g$ was reduced by up to 20%, though for $S/N \gtrsim 50$, the difference was negligible.

We incorporated photometry as follows. We took magnitudes (and the associated uncertainties) from eight bands in APASS DR9 (Henden & Munari 2014; Henden et al. 2016), 2MASS (Cutri et al. 2003; Skrutskie et al. 2006) and *WISE* (Wright et al. 2010): *g, r, i, J, H, K, W1* (3.4 $\mu$m), and *W2* (4.6 $\mu$m). From these, we constructed seven colors: *g-r, r-i, i-J, J-H, H-K, K-W1*, and *W2-W1*. For each reference object, we added its seven colors as "pixels" to its spectrum: the color as the "flux" and the uncertainty as the "error bar." Using colors restricted us to the set of reference objects with APASS, 2MASS, and *WISE* magnitudes: 8472 of the 9173.

We emphasize that *The Cannon* builds a model of spectra but is agnostic to whether the value of a "spectral pixel" is a true flux value or simply another observed property of the star that is sensitive to the labels of interest.

---

[1] Because $\alpha$-enhanced stars have a different line-of-sight velocity distribution, velocity and [$\alpha$/M] may well be correlated.

### 3.2. Results from Cross-validation

We evaluate the accuracy and precision of the model using a "leave-$\frac{1}{8}$-out" cross-validation test. We split the 8472 reference objects into eight groups, by assigning each one a random integer between 0 and 7. We leave out each group in turn, and train a model on the remaining seven groups. We then apply that model to infer new labels for the group that was left out. We use the results of the cross-validation to determine which objects are appropriate reference objects for training the model: we excise objects whose difference from the training value is greater than four times the scatter in that label, leaving 8125 objects. We train the model on these 8125 objects and re-run the cross-validation. We also use the model to infer labels for the 347 objects excised from the training, in order to properly account for all of the objects in the following error analysis.

At the end of this process, each of the 8472 objects has a new set of labels determined by *The Cannon*, from a model that was *not* trained using that object. Figure 1 shows the comparison of these Cannon-inferred "test" labels with the reference labels used in training, for high-S/N objects; there is a significant decrease in scatter compared to the corresponding figure in Paper 1 (Fig. 6). For objects with S/N > 100, the labels are consistent with the APOGEE training values to within 53 K in $T_{\text{eff}}$, 0.11 dex in log g, 0.05 dex in [M/H], 0.06 dex in [C/M], 0.09 dex in [N/M], 0.03 dex in [$\alpha$/M], and 0.04 mag in $A_k$. These are comparable to, or within, the stated ASPCAP uncertainties (Holtzman et al. 2015).

Figure 2 shows the scatter in different bins of S/N (where S/N is the median value of formal S/N across all pixels in the spectrum) in all of the labels except for $A_k$ (which is primarily determined from the additional photometric pixels, not taken into account in determining the S/N). By construction, as a result of this data-driven label transfer, there is significant improvement in agreement with the APOGEE values over those from the LAMOST pipeline.

### 4. MEASURING CARBON AND NITROGEN FROM LAMOST SPECTRA

A key strength of *The Cannon* is the physical interpretability of the spectral model. Since an independent model is fit at every pixel of the spectrum, each pixel has a set of model coefficients as well as a model scatter term. The leading (linear) coefficient in each label can be thought of as a proxy for the sensitivity of each pixel to a particular label; thus, each label has a wavelength-dependent indicator of which spectral regions are most informative. Each pixel also has a model scatter term; this is the intrinsic variance in the model at that pixel, as distinct from the observational variance. In other words, it is the expected deviation from the model at that particular pixel, in the limit of a perfect measurement.

These are shown in Figure 3, in the part of the spectrum found by *The Cannon* to be most predictive of labels (the blue end, $\sim 4000 - 5800$ Å), together with the scatter in the model. These linear coefficients are the first derivatives of the model at a set of fiducial stellar parameters; in this case, we pivot the model around the mean value in each label across the training set: $T_{\rm eff} = 4687\,{\rm K}$, $\log g = 2.83\,{\rm dex}$, $[M/H] = -0.19\,{\rm dex}$, $[\alpha/M] = -0.01\,{\rm dex}$, $[C/M] = 0.10\,{\rm dex}$, $[N/M] = 0.09\,{\rm dex}$, and $A_{\rm k} = 0.09\,{\rm mag}$. To facilitate comparison, each derivative has been scaled by the approximate error in the corresponding label: $\delta T_{\rm eff} \sim 91.5\,{\rm K}$, $\delta \log g \sim 0.11\,{\rm dex}$, $\delta[M/H] \sim 0.05\,{\rm dex}$, $\delta[\alpha/M] \sim 0.05\,{\rm dex}$, $\delta[C/M] \sim 0.04\,{\rm dex}$, and $\delta[N/M] \sim 0.07\,{\rm dex}$ (Holtzman et al. 2015).

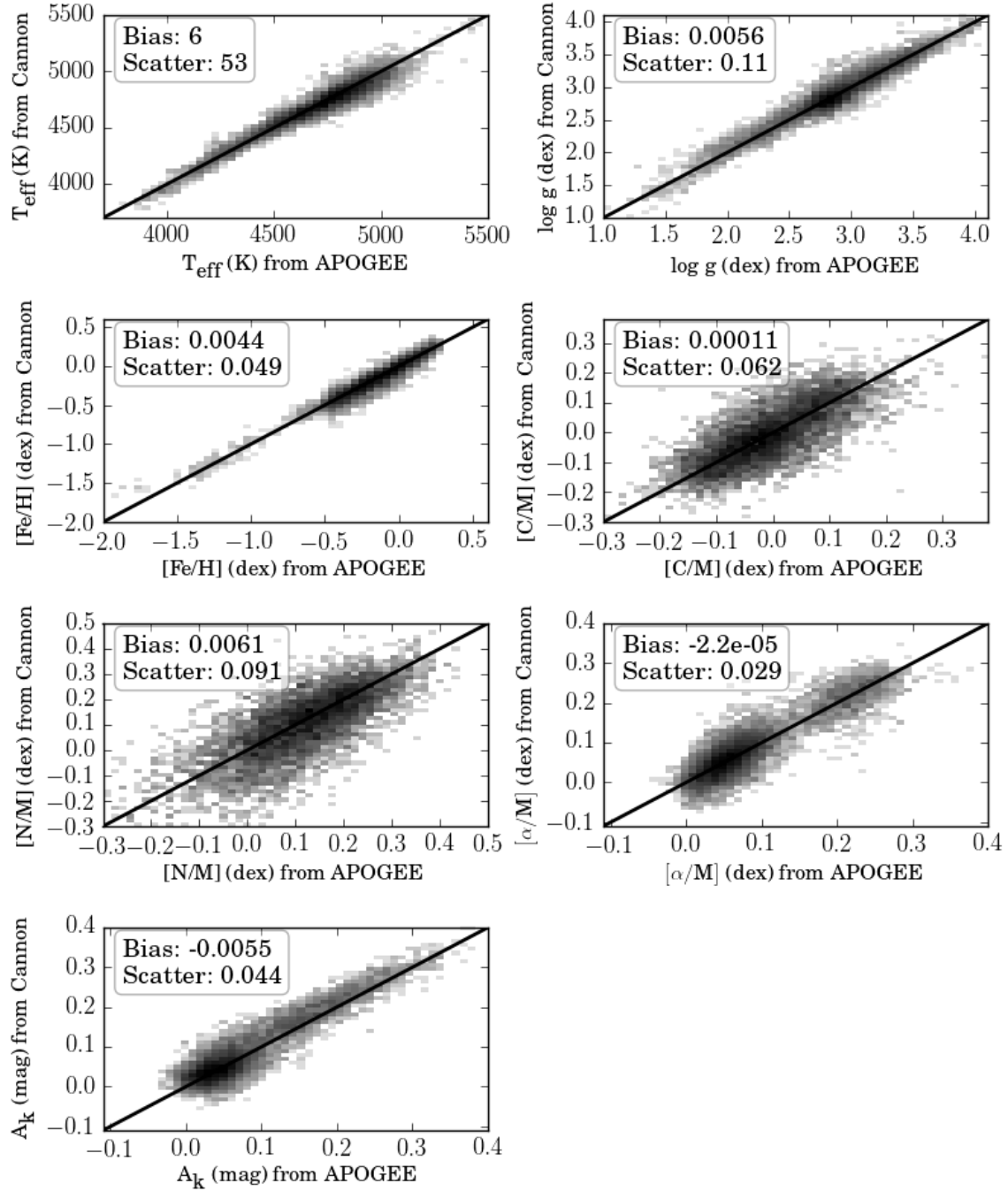

**Figure 1.** Results from cross-validation of *The Cannon*'s label transfer from APOGEE to LAMOST, for spectra with S/N > 100. The model was trained on APOGEE labels ($x$-axis) and used to infer new labels ($y$-axis) from the corresponding LAMOST spectra. The improvement in precision over the results in Paper 1 (see Fig. 6) reflects changes we made to the model, described in Section 3. The low bias and scatter in [C/M] and [N/M] demonstrates that these abundances can in fact be measured from low-resolution LAMOST spectra.

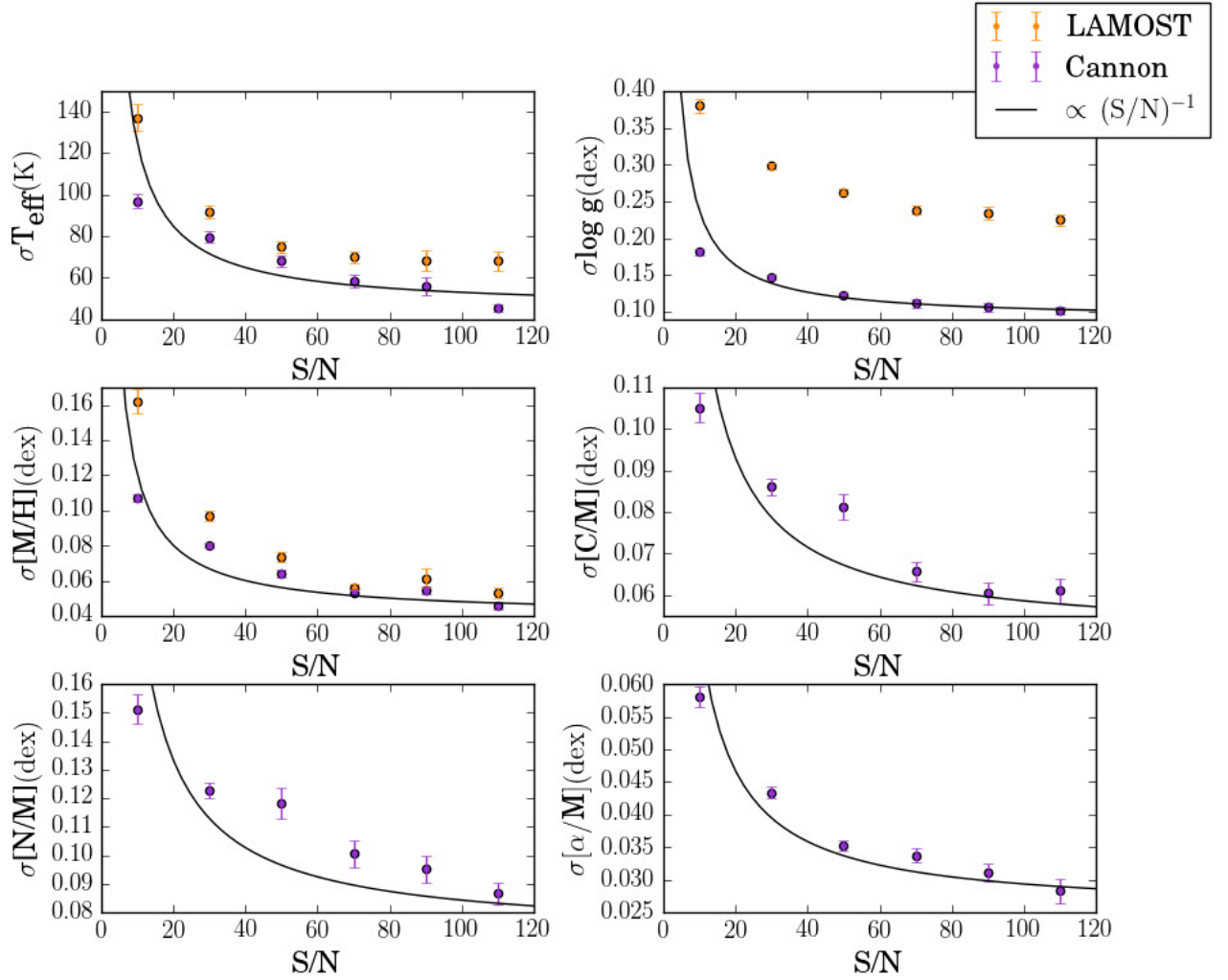

**Figure 2.** The S/N-dependence of the scatter between APOGEE DR12 labels and the corresponding labels measured from LAMOST spectra by *The Cannon* (purple points) and the LAMOST pipeline (yellow points), for 8472 objects. By construction, the labels measured by *The Cannon* are in closer agreement with the APOGEE values for $T_{\rm eff}$, $\log g$, and [M/H], and the model behaves well with decreasing S/N. Note that we are using our own formal measurement of S/N, not the reported LAMOST value. The $(S/N)^{-1}$ curves were shifted vertically by eye.

The new labels in this work are [C/M] and [N/M], so we should demonstrate that *The Cannon* measures these values from physically plausible spectral signatures. Ting et al. (2017) (submitted to ApJ) quantified how the precision with which various abundances (including [C/H] and [N/H]) can be measured varies as a function of survey resolution. They showed that low-resolution ($R < 10,000$) spectra, such as those from LAMOST, have the same theoretically achievable uncertainties per stellar label as medium-resolution ($10,000 < R < 50,000$) spectra, under the following assumptions: equal exposure time (so that a low-resolution spectrum has higher S/N per resolution element), an equal number of detector pixels (so that low-resolution spectra have more extensive wavelength coverage), and a constant sampling per resolution element. These predictions are based in part on gradient spectra calculated using Kurucz models (Kurucz 1970, 1979, 1993, 2005; Kurucz & Avrett 1981), which are assumed to be perfect. Of course, this aspect does not pertain to data-driven models (see the discussion in Ting et al. (2016a)).

To make a direct comparison between the Cannon model and theoretical predictions, we calculate gradient spectra and compare them to the gradient spectra calculated from Kurucz models by Ting et al. (2016a). Gradient spectra are a quantification of how much the flux at a given wavelength changes with changes to a given label: in other words, it characterizes the sensitivity or information content of each wavelength for a given label. Following Equation 2 in Ting et al. (2017), gradient spectra are calculated as follows.

$$\nabla_\ell f_{\rm model}(\lambda, \ell_i) = \frac{f_{\rm model}(\lambda, \ell_i + \Delta \ell_i) - f_{\rm model}(\lambda, \ell_i)}{\Delta \ell_i} \quad (1)$$

where $f_{\rm model}(\lambda, \ell_i)$ represents a model spectrum across wavelengths $\lambda$, generated using a set of labels $\ell_i$. A fractional change in a particular label $\Delta \ell_i$ results in a fractional change in the spectrum $\nabla_\ell f_{\rm model}$ at each wavelength $\lambda$. In other words, to study sensitivity of a spectral region to a particular label, one changes the value in that label and calculates the fractional change in the flux in that region.

We have two sets of model spectra: one from the Cannon model as described in Section 3, and one from the Kurucz models, in both cases generated using labels for a solar-metallicity K-giant ($T_{\rm eff} = 4800$, $\log g = 3.5$). We use each of these models to calculate a gradient spectrum for [C/M] by varying [C/M] by 0.2 dex, and a gradient spectrum for [N/M] by varying [N/M] by 0.2 dex. For a more direct comparison, we normalize the theoretical gradient spectra the same way as the LAMOST spectra.

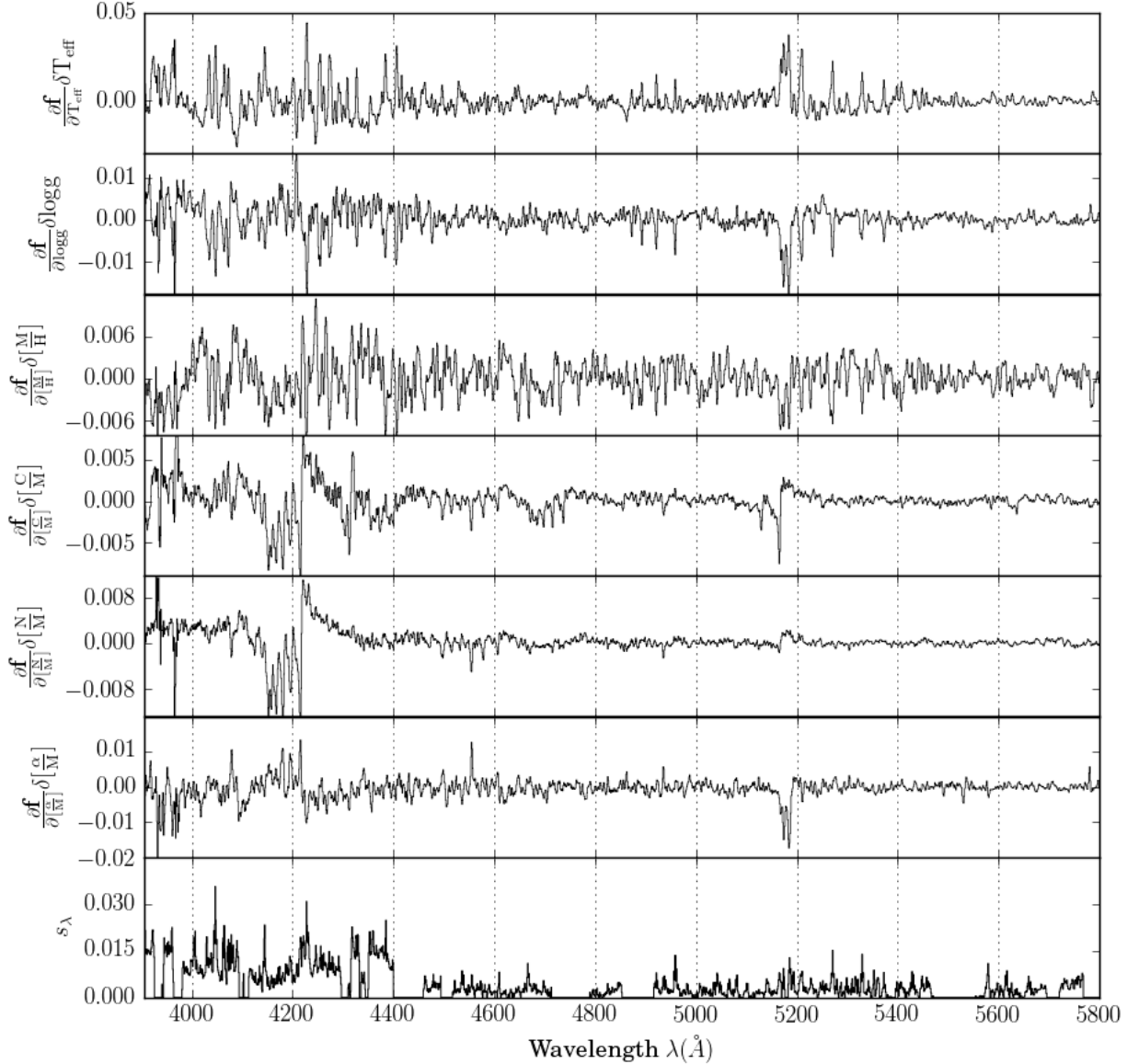

**Figure 3.** Leading (linear) coefficients and scatter from the best-fit spectral model, determined by *The Cannon* using the 8125 reference objects. The leading coefficients can be thought of as a proxy for the sensitivity of each pixel in the spectrum to each of the labels; to facilitate comparison, each has been scaled by the approximate error in that label (Holtzman et al. 2015). The scatter term $s_\lambda$ is the variance intrinsic to the model; it quantifies each pixel's expected deviations from the model in the limit of a perfect measurement. We display the model coefficients and scatter blueward of 5800 Å because this was found by *The Cannon* to be the region with the most sensitive features.

Figure 4 shows the Cannon model gradient spectra overlaid with the Kurucz model gradient spectra from Ting et al. (2017), for the CN and CH molecular features in the wavelength range 4100-4400 Å (see Martell et al. 2008). The panel on the left shows the gradient spectra for carbon, and the panel on the right shows the gradient spectra for nitrogen. Because the theoretical gradient spectra (red) were generated purely from physical models, and the Cannon gradient spectra (black) represent an entirely data-driven relationship between these wavelength regions and abundances from APOGEE, the qualitative similarity between them is gratifying. There are clearly some quantitative differences, but we simply seek to demonstrate here that the signatures of carbon and nitrogen from the data-driven Cannon model come from astrophysically sensible wavelength regions, such as the 4215 Å CN band. Furthermore, the differences between the two panels demonstrates that we are measuring each element from distinct features, not simply correlations between the two (e.g. the carbon-sensitive CH (G) band, not present in the nitrogen signature). The fact that [C/M] and [N/M] share regions of sensitivity does not mean that they are degenerate; they may be covariant, but can still be independently measured when fit for simultaneously (see Ting et al. 2017, submitted).

## 5. FROM [C/M] AND [N/M] TO MASS AND AGE

After cross-validating the model, we use it to infer labels for the same 454,180 objects described in Paper 1. These were identified as giants using their LAMOST labels. For a detailed description of this selection procedure, we direct the reader to Section 4.2 of that paper. In short, the LAMOST values of $T_{\text{eff}}$ and $\log g$ enable a reliable dwarf-giant separation.

To transform [C/M] and [N/M] into mass and age, we use the formulas characterized by the coefficients in Tables A2 and A3 of Martig et al. (2016), which are in turn based on asteroseismic mass measurements for stars with [C/M] and [N/M] measurements. These relations are only applicable within a certain range of label values, restricting the number of objects for which we can infer masses and ages via their [C/M] and [N/M]. Although we infer [C/M] and [N/M] for the full set of 454,180 test objects described in Paper 1, we apply the following cuts (following Martig et al. 2016), which leaves 230,901 objects suitable for the formula.

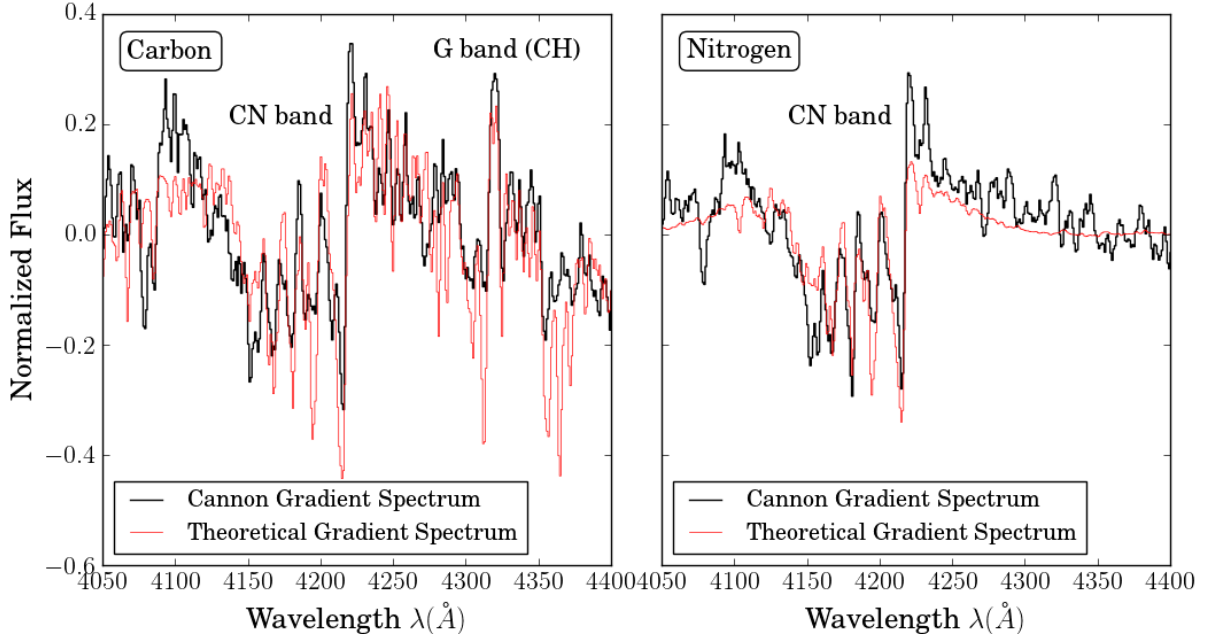

**Figure 4.** Gradient spectra for carbon (left panel) and nitrogen (right panel) calculated using two different models: theoretical Kurucz models (red line) and the Cannon model (black line). All model spectra were generated using K-giant ($T_{eff}$=4750, log g=2.5) and solar-metallicity values, stepping over 0.2 dex in [C/M] and [N/M]. For a better comparison, the Kurucz gradient spectra were normalized the same way as the LAMOST spectra, as described in Section 3. The qualitative similarity between the black and red lines demonstrates that the Cannon measurements of [C/M] and [N/M] are coming from astrophysically sensible spectral regions, e.g. the 4215 Å CN band. Furthermore, the difference between the left and right panels demonstrates that [C/M] and [N/M] are being measured independently, not just via correlations: for example, the nitrogen signature does not include any changes in the G (CH) band.

$$\begin{cases} -0.8 < [\text{M/H}] < 0.25 \\ 4000 < T_{eff} < 5000 \\ 1.8 < \log g < 3.3 \\ -0.25 < [\text{C/M}] < 0.15 \\ -0.1 < [\text{N/M}] < 0.45 \\ -0.05 < [\alpha/\text{M}] < 0.3 \\ -0.1 < [(\text{C}+\text{N})/\text{M}] < 0.15 \\ -0.6 < [\text{C/N}] < 0.2 \end{cases}$$

These cuts set the primary restriction on the size of our mass and age catalog; for example, although the LAMOST data contain a large population of low-metallicity outer disk stars, we cannot estimate masses and ages for those objects.

The estimated masses and ages have uncertainties that come from the intrinsic scatter of the relation (Martig et al. 2016) and the individual stellar label uncertainties.

To estimate the latter, we sample from each star's label *pdf* 100 times, approximating each distribution as a Gaussian with a standard deviation equal to the scatter in that label. That scatter is a function of S/N (see Figure 2), so for each object we take the standard deviation of its Gaussian spread to be the scatter associated with that S/N bin. Thus, each object has a distribution in mass and age. We take the mass and age value to be the median of that distribution, and estimate the uncertainty using the half-width of the central 68$^{\text{th}}$ percentile.

This procedure does not account for the errors in the training labels or the scatter in the Martig et al. (2016) relation. There are also systematic errors from the functional form of the age relation in Martig et al. (2016) that are not taken into account here. In other words, our mass and age estimates are within the framework of the functional forms in Martig et al. (2016), but as we provide the [C/M] and [N/M] measurements separately, others could redo the [C/M] to [N/M] calibration using our values. Note that these are all distinct from, and in addition to, the formal error from the Cannon model fit.

We provide a catalog of all our inferred labels, including mass and age; an excerpt is shown in Table 1. We provide the best-fit values for each label, including [C/M] and [N/M] for all 454,364 objects, and mass and age for the 230,901 of those that fall within the label space of Martig et al. (2016). We also provide the associated formal errors from the covariance matrix of the fit. Note that these formal errors are certainly an underestimate. There are also contributions from the uncertainties in the reference labels (not taken into account here) and the discreteness of the reference set (estimated in Section 4.2 of Paper 1 to be very small).

A more accurate estimate of the uncertainties is provided by the spread in the cross-validation (see Section 3.2, Figure 1, and Figure 2). For the convenience of those who wish to use our catalog, we provide an estimate of these uncertainties by fitting a quadratic function to the purple (Cannon) points in Figure 2. Thus, for each object, we use the S/N of its spectrum to estimate the uncertainty. These uncertainty values are labeled as the "scatter" in the table.

We also provide the formal S/N for the spectrum of each object and the reduced chi-squared of the fit, which is the chi-squared divided by the approximate number of pixels in each spectrum ($\sim 1800$). Note that the S/N and the reduced chi-squared are both low by roughly a factor of three; see the discussion in Section 4.1 of Paper 1. Furthermore, note that the values of $T_{\text{eff}}$, $\log g$, [M/H], and [$\alpha$/M] will not be identical to their corresponding values in Paper 1 for several reasons: they are on the uncalibrated APOGEE label scale, and there have been various changes in our procedure (masking 50% of the spectrum, the inclusion of photometry, fitting for additional labels).

Because of the addition of photometry and the masking, we believe that the values in this table are more accurate and precise than the values in the corresponding table from Paper 1. Therefore, we encourage readers to use the table in this paper for values of LAMOST labels.

**Table 1.** Excerpt from the online table of stellar labels ($T_{\rm eff}$, $\log g$, [M/H], [C/M], [N/M], [$\alpha$/M], $A_k$, mass, and age) for 454,364 stars, inferred by *The Cannon*. Column 1 is the LAMOST ID of the object, Columns 2-3 are the position of the object in RA and Dec (J2000), Columns 4-10 are the labels from *The Cannon*, Columns 11-17 are the formal errors on the Cannon-inferred labels from the covariance matrix in the model fit, Columns 18-23 are the estimated uncertainties based on the scatter in the cross-validation, as described in Section 5, Columns 24-25 are the derived masses and ages, Columns 26-27 are the estimated uncertainties on mass and age, and Columns 28-29 are the S/N of the spectrum and the reduced $\chi^2$ of the model fit. Note that the reduced $\chi^2$ values are low by a factor of $\sim 3$ because the random component of the errors in the LAMOST spectra is overestimated.

| LAMOST ID | RA | Dec | $T_{\rm eff}$ | $\log g$ | [M/H] | [C/M] | [N/M] | [$\alpha$/M] | $A_k$ |
|---|---|---|---|---|---|---|---|---|---|
| | (deg) | (deg) | (K) | (dex) | (dex) | (dex) | (dex) | (dex) | mag |
| spec-55859-F5902_sp01-034.fits | 331.91682 | -1.78259 | 4793 | 3.22 | -0.507 | 0.0645 | -0.0242 | 0.228 | 0.054 |
| spec-55859-F5902_sp03-209.fits | 331.13586 | 0.85349 | 4620 | 2.88 | -0.347 | 0.0984 | 0.107 | 0.22 | 0.0131 |
| spec-55859-F5902_sp06-160.fits | 334.26706 | -0.15868 | 4240 | 2.23 | -0.293 | 0.0734 | 0.102 | 0.208 | 0.148 |
| spec-55859-F5902_sp08-146.fits | 333.40638 | -0.3965 | 4894 | 3.29 | -0.337 | -0.0221 | -0.0243 | 0.212 | 0.0293 |

**Table 1.** continued: Formal Errors

| LAMOST ID | $\sigma(T_{\rm eff})$ | $\sigma(\log g)$ | $\sigma$([M/H]) | $\sigma$([C/M]) | $\sigma$([N/M]) | $\sigma$([$\alpha$/M]) | $\sigma(A_k)$ |
|---|---|---|---|---|---|---|---|
| | (K) | (dex) | (dex) | (dex) | (dex) | (dex) | (mag) |
| spec-55859-F5902_sp01-034.fits | 57 | 0.1 | 0.052 | 0.0657 | 0.0945 | 0.026 | 0.019 |
| spec-55859-F5902_sp03-209.fits | 8 | 0.02 | 0.009 | 0.0108 | 0.017 | 0.01 | 0.0072 |
| spec-55859-F5902_sp06-160.fits | 8 | 0.02 | 0.01 | 0.0089 | 0.012 | 0.007 | 0.008 |
| spec-55859-F5902_sp08-146.fits | 71 | 0.13 | 0.069 | 0.0651 | 0.0761 | 0.038 | 0.0204 |

**Table 1.** continued: Estimated Error (Scatter)

| LAMOST ID | $s(T_{\rm eff})$ | $s(\log g)$ | $s$([M/H]) | $s$([C/M]) | $s$([N/M]) | $s$([$\alpha$/M]) |
|---|---|---|---|---|---|---|
| | (K) | (dex) | (dex) | (dex) | (dex) | (dex) |
| spec-55859-F5902_sp01-034.fits | 78 | 0.14 | 0.078 | 0.0851 | 0.1232 | 0.043 |
| spec-55859-F5902_sp03-209.fits | 43 | 0.1 | 0.046 | 0.0615 | 0.087 | 0.03 |
| spec-55859-F5902_sp06-160.fits | 43 | 0.1 | 0.046 | 0.0615 | 0.087 | 0.03 |
| spec-55859-F5902_sp08-146.fits | 87 | 0.16 | 0.092 | 0.0955 | 0.1353 | 0.049 |

**Table 1.** continued

| LAMOST ID | Mass | log(Age) | $\sigma$(Mass) | $\sigma$(log(Age)) | SNR | Red. $\chi^2$ |
|---|---|---|---|---|---|---|
| | ($M_\odot$) | dex | ($M_\odot$) | (dex) | | |
| spec-55859-F5902_sp01-034.fits | 0.79 | 1.0 | 0.39 | 0.4 | 33.7 | 0.44 |
| spec-55859-F5902_sp03-209.fits | 1.0 | 0.85 | 0.1 | 0.14 | 169.0 | 1.7 |
| spec-55859-F5902_sp06-160.fits | 1.4 | 0.64 | 0.6 | 0.45 | 130.0 | 1.2 |
| spec-55859-F5902_sp08-146.fits | 1.3 | 0.64 | 0.5 | 0.48 | 19.9 | 0.51 |

5.1. *Astrophysical Verification of Inferred Ages*

We now investigate whether our inferred age values seem astrophysically plausible. Figure 5 shows the ([M/H], [α/M]) plane color-coded by age for the 42,420 objects with S/N > 80. The color of each bin reflects the mean age of stars in that range of [M/H] and [α/M], weighted by the estimated uncertainty in the age measurement. We do not show bins with fewer than 20 objects. We see an astrophysically sensible age gradient with changing abundances, from the young, low-[α/M] sequence to the old, high-[α/M] sequence. This is qualitatively very similar to the gradient seen from small high-resolution datasets of main-sequence turn-off stars in the solar neighborhood (e.g. Haywood et al. 2013).

Furthermore, as Figure 6 shows, our masses and ages (for the reference objects) are in remarkable agreement with the masses and ages from the Ness et al. (2016) catalog, determined via a rather different approach. In that approach ($x$ axis) masses were measured directly from APOGEE spectra ($R \sim 22,500$) and ages were estimated via isochrone fitting with no connection to carbon or nitrogen. In our approach ($y$ axis) we made use of the relations in Martig et al. (2016), which have no connection to spectra. It is true that the Martig et al. (2016) relation and the Ness et al. (2016) catalog were both calibrated, or trained, on the same sample of stars: the 1639-object APOKASC sample. However, here we are showing the 6125 objects in common between LAMOST and the Ness et al. (2016) sample, which has almost no overlap with APOKASC. The agreement between the two approaches outside of the set used for calibration supports the plausibility of our estimates.

Figure 6 also includes the marginalized histogram of mass and age respectively. The full span of mass is 0.6 dex and the full span of age is 1.5 dex. The errors in each label are 7.5 smaller than this. However, the bulk of the objects fall within a narrower distribution. Clearly, this allows only broad sorting and not precision mass measurements or age dating.

Finally, Figure 7 shows the spatial distribution of our sample, and its expansion over that from 70,000 stars in APOGEE (Ness et al. 2016). As expected, younger stars are concentrated toward the disk mid-plane, and older stars extend to a larger scale height away from the disk and into the bulge and halo.

## 6. DISCUSSION

Using a data-driven approach to spectral modeling, and fitting for all labels simultaneously, we find that we can measure accurate and precise carbon and nitrogen abundances from low-resolution ($R \sim 1800$) LAMOST spectra. For post dredge-up giants, as in the sample from Martig et al. (2016), these [C/M] and [N/M] measurements enable mass and age estimates across the sky, to 0.08 dex in mass and 0.2 dex in age. With this new set of ages, we have a very different spatial sampling than APOGEE: we have essentially tied in-the-disk and off-the-disk ages onto the same scale, as LAMOST has a much better sampling of the thick disk than APOGEE.

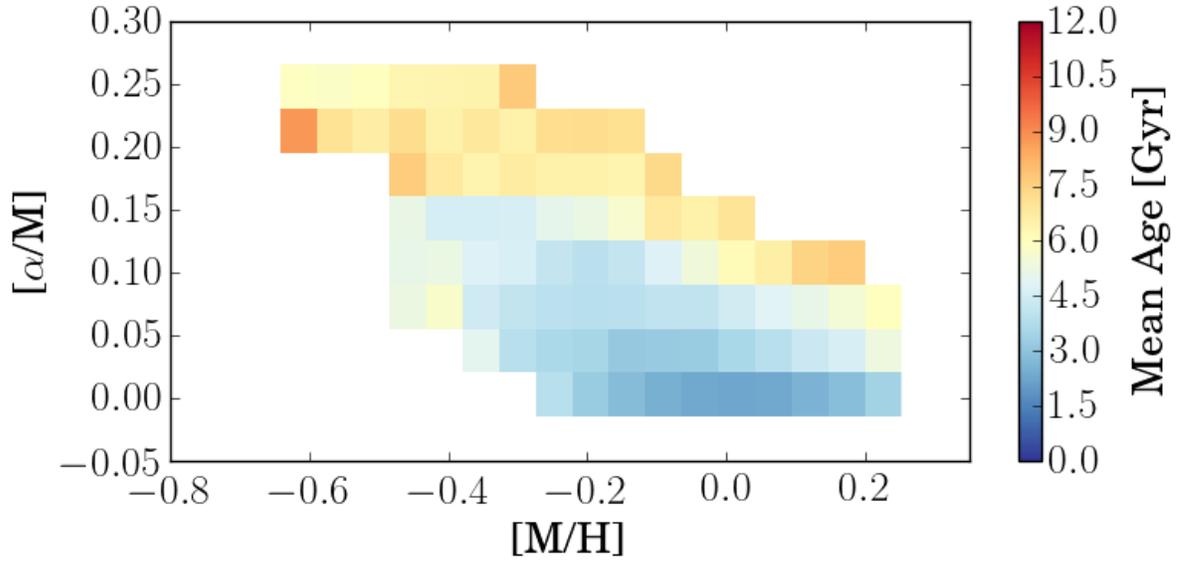

**Figure 5.** Cannon age estimates for LAMOST giants: the [α/M]-[M/H] plane color-coded by the mean age (weighted by the estimated age uncertainties) in each bin. We excised objects with S/N < 80 (leaving 42,420 objects) and only show bins with > 20 objects. The gradient from young α-poor to old α-rich stars is astrophysically very plausible.

The success of our data-driven approach in extracting information on [C/M] and [N/M] from blended regions (see Figure 4) holds promise for a natural extension of this work to measuring additional individual element abundances. *The Cannon* has already been successful at measuring individual abundances from APOGEE spectra, in part because the model is not restricted to unblended element windows (Casey et al. 2016; Hogg et al. 2016; Ness et al. 2016). Indeed, Ting et al. (2017) predict using theoretical models that spectra of comparable resolution to LAMOST should not only contain sufficient information to precisely constrain [C/M] and [N/M], but also a large suite of other individual element abundances, such as aluminum, calcium, manganese, and nickel.

For the purpose of [C/M] and [N/M] measurement in this work, it was helpful to apply broad masks to the spectra, to fully remove telluric and interstellar absorption features. Depending on which spectral regions encode information on [X/H], however, fitting for these additional labels would likely require more precise masking in order to avoid removing important signatures.

Finally, at nearly identical values of $\{T_{\rm eff}, \log g, [{\rm Fe/H}]\}$ on the giant branch, mass or age is highly predictive of luminosity. Age constraints to 0.2 dex could therefore be useful for improving estimates of stellar luminosity. Since luminosity depends on distance at a given $T_{\rm eff}$, [Fe/H], and $\log g$, these age constraints also help to estimate distance. How well this works in practice would need to be explored. However, as a first pass, we estimate the improvement using a set of Padova isochrones for stars of indistinguishable $T_{\rm eff}$, $\log g$, and [Fe/H] (that is, within a standard deviation in $\log g$ and $T_{\rm eff}$, for [Fe/H] = 0). This suggests that an age estimate to 0.2 dex would constrain luminosity to 0.2 mag, or 10% in distance.

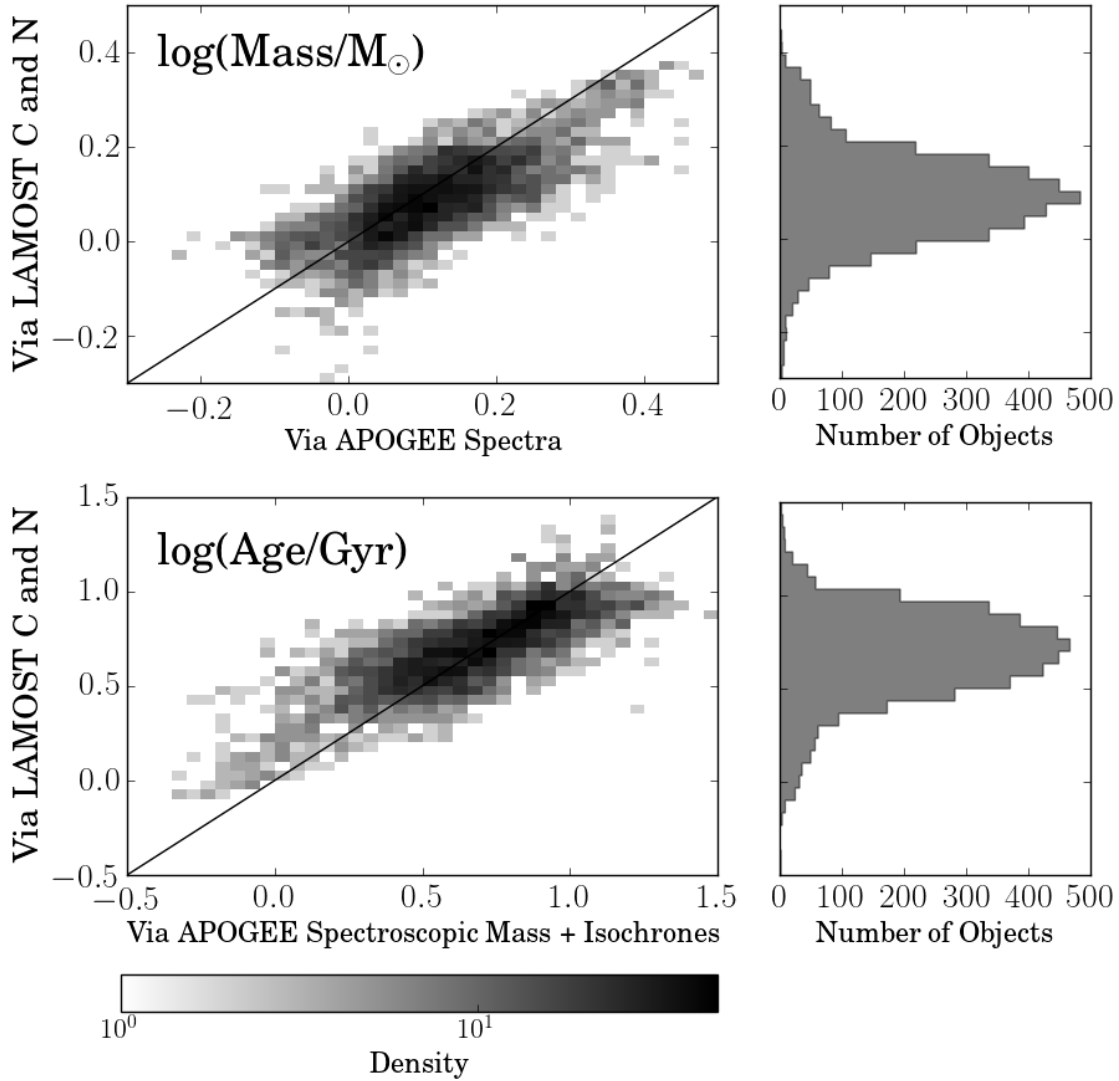

**Figure 6.** For the 6215 objects in common between our LAMOST sample and the APOGEE mass and age catalog in Ness et al. (2016): comparison between our estimates (inferred via [C/M] and [N/M] abundances) and the Ness et al. (2016) mass and age estimates (inferred via spectroscopic mass measurements and isochrone fitting). The top panel shows the comparison for mass and the bottom panel shows the comparison for age. The histograms on the right show the distribution of mass and age. The agreement with the Ness et al. (2016) values despite the two very different approaches supports the plausibility of our measurements.

The code used to produce the results described in this paper was written in Python and is available online in an open-source repository.[2]

---

[2] www.github.com/annayqho/TheCannon

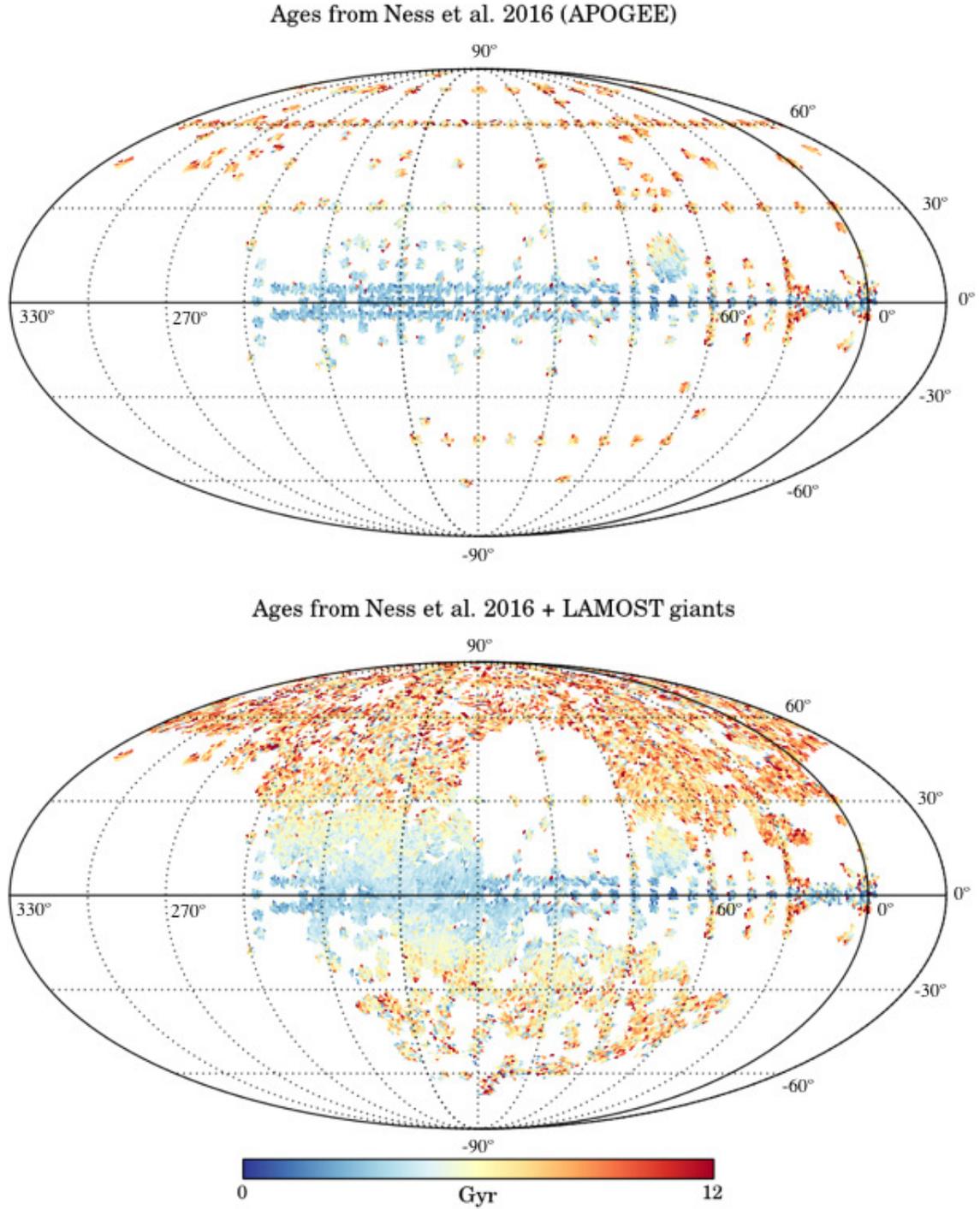

**Figure 7.** Distribution on the sky (in Galactic coordinates) of stars with age measurements: the top panel shows the sample from Ness et al. (2016) (∼70,000 objects) and the bottom panel overlays these values with 230,901 ages inferred via [C/M] and [N/M] by *The Cannon* from the LAMOST spectra. The much more extensive area coverage of the LAMOST data is immediately apparent.

We are grateful to Evan Kirby (Caltech) for his careful reading and constructive feedback on the manuscript. It is furthermore a pleasure to thank Jon Bird (Vanderbilt), Sven Buder (MPIA), Andy Casey (IoA Cambridge), Christina Eilers (MPIA), Jim Fuller (Caltech), Lynne Hillenbrand (Caltech), Sarah Martell (UNSW), and Marie Martig (MPIA) for helpful discussions and assistance. A.Y.Q.H. is grateful to the community at the MPIA for their support and hospitality during the period in which much of this work was performed. H.W.R. acknowledges support of the Miller Institute at UC Berkeley through a visiting professorship during the completion of this work. Finally, we would like to thank the anonymous referee whose detailed comments greatly improved the clarity and precision of the paper.

A.Y.Q.H. was supported by a National Science Foundation Graduate Research Fellowship under Grant No. DGE1144469. M.K.N. and H.W.R. have received funding for this research from the European Research Council under the European Union's Seventh Framework Programme (FP 7) ERC Grant Agreement n. [321035]. D.W.H. was partially supported by the NSF (grant IIS-1124794), NASA (grant NNX08AJ48G), and the Moore-Sloan Data Science Environment at NYU. C.L. acknowledges the Strategic Priority Research Program "The Emergence of Cosmological Structures" of the Chinese Academy of Sciences, Grant No. XDB09000000, the National Key Basic Research Program of China 2014CB845700, and the National Natural Science Foundation of China (NSFC) grants No. 11373032 and 11333003.

Guoshoujing Telescope (the Large Sky Area Multi-Object Fiber Spectroscopic Telescope LAMOST) is a National Major Scientific Project built by the Chinese Academy of Sciences. Funding for the project has been provided by the National Development and Reform Commission. LAMOST is operated and managed by the National Astronomical Observatories, Chinese Academy of Sciences.

Funding for the Sloan Digital Sky Survey IV has been provided by the Alfred P. Sloan Foundation, the U.S. Department of Energy Office of Science, and the Participating Institutions. SDSS-IV acknowledges support and resources from the Center for High-Performance Computing at the University of Utah. The SDSS web site is www.sdss.org.

SDSS-IV is managed by the Astrophysical Research Consortium for the Participating Institutions of the SDSS Collaboration including the Brazilian Participation Group, the Carnegie Institution for Science, Carnegie Mellon University, the Chilean Participation Group, the French Participation Group, Harvard-Smithsonian Center for Astrophysics, Instituto de Astrofísica de Canarias, The Johns Hopkins University, Kavli Institute for the Physics and Mathematics of the Universe (IPMU) / University of Tokyo, Lawrence Berkeley National Laboratory, Leibniz Institut für Astrophysik Potsdam (AIP), Max-Planck-Institut für Astronomie (MPIA Heidelberg), Max-Planck-Institut für Astrophysik (MPA Garching), Max-Planck-Institut für Extraterrestrische Physik (MPE), National Astronomical Observatory of China, New Mexico State University, New York University, University of Notre Dame, Obser-

22
vatário Nacional / MCTI, The Ohio State University, Pennsylvania State University, Shanghai Astronomical Observatory, United Kingdom Participation Group, Universidad Nacional Autónoma de México, University of Arizona, University of Colorado Boulder, University of Oxford, University of Portsmouth, University of Utah, University of Virginia, University of Washington, University of Wisconsin, Vanderbilt University, and Yale University.

This research has made use of the APASS database, located at the AAVSO web site. Funding for APASS has been provided by the Robert Martin Ayers Sciences Fund.

This publication makes use of data products from the Two Micron All Sky Survey, which is a joint project of the University of Massachusetts and the Infrared Processing and Analysis Center/California Institute of Technology, funded by the National Aeronautics and Space Administration and the National Science Foundation.

*Facilities:*

*Facility:* Sloan(APOGEE spectrograph),

*Facility:* LAMOST,

*Facility:* WISE,

*Facility:* AAVSO